\documentclass[conference]{IEEEtran}
\IEEEoverridecommandlockouts
\usepackage{cite}
\usepackage{amsmath,amssymb,amsfonts}
\usepackage{algorithmic}
\usepackage{graphicx}
\usepackage{textcomp}
\usepackage{xcolor}
\usepackage{orcidlink}
\usepackage{lipsum}
\usepackage[utf8]{inputenc} 
\usepackage[T1]{fontenc}    
\usepackage{hyperref}       
\usepackage{url}            
\usepackage{booktabs}       
\usepackage{nicefrac}       
\usepackage{microtype}      
\usepackage{float}

\usepackage{fancyhdr} 
\def\BibTeX{{\rm B\kern-.05em{\sc i\kern-.025em b}\kern-.08em
    T\kern-.1667em\lower.7ex\hbox{E}\kern-.125emX}}
\begin{document}

\title{HiFi-HARP: A High-Fidelity 7th-Order Ambisonic Room Impulse Response Dataset}


\author{
\IEEEauthorblockN{
\IEEEauthorblockN{Shivam Saini\orcidlink{0009-0002-3981-2260}
and Jürgen Peissig}}
\IEEEauthorblockA{Institut für Kommunikationstechnik \\
Leibniz University, Hannover\\}
\thanks{Samples available at: \url{https://github.com/whojavumusic/hifi-harp}}
}

\maketitle

\begin{abstract}
  We introduce HiFi-HARP, a large-scale dataset of 7th-order Higher-Order Ambisonic Room Impulse Responses (HOA-RIRs) consisting of more than 100,000 RIRs generated via a hybrid acoustic simulation in realistic indoor scenes. HiFi-HARP combines geometrically complex, furnished room models from the 3D-FRONT repository with a hybrid simulation pipeline: low-frequency wave-based simulation (finite-difference time-domain) up to 900 Hz is used, while high frequencies above 900 Hz are simulated using a ray-tracing approach. The combined raw RIRs are encoded into the spherical-harmonic domain (AmbiX ACN) for direct auralization. Our dataset extends prior work by providing 7th-order Ambisonic RIRs that combine wave-theoretic accuracy with realistic room content. We detail the generation pipeline (scene and material selection, array design, hybrid simulation, ambisonic encoding) and provide dataset statistics (room volumes, RT60 distributions, absorption properties). A comparison table highlights the novelty of HiFi-HARP relative to existing RIR collections. Finally, we outline potential benchmarks such as FOA-to-HOA upsampling, source localization, and dereverberation. We discuss machine learning use cases (spatial audio rendering, acoustic parameter estimation) and limitations (e.g., simulation approximations, static scenes). Overall, HiFi-HARP offers a rich resource for developing spatial audio and acoustics algorithms in complex environments.
\end{abstract}

\section{Introduction}
Higher-Order Ambisonics (HOA) is a powerful framework for spatial audio in applications like VR/AR, offering high quality reproduction \cite{hoainvest}. However, the development of advanced HOA-based algorithms such as source localization and sound field synthesis is hampered by the lack of large-scale high-fidelity Room Impulse Response (RIR) datasets.

Existing large-scale synthetic datasets have key limitations: SoundSpaces offers millions of RIRs but is limited to First-Order Ambisonics (FOA) and purely geometric simulation \cite{chen20soundspaces}; the hybrid-acoustic GWA dataset is single-channel only \cite{GWA}; and the 7th-order HARP dataset uses the image-source method in simple shoebox rooms \cite{saini2025harp}. While measured datasets exist \cite{murphy2010openair, politis2022tau}, they are limited in scale and variety. This landscape reveals a clear need for a large-scale HOA dataset that combines high-fidelity hybrid simulation with realistic room geometries.

\section{Related Work }
\label{sec:related}

\textbf{HOA and Ambisonic datasets:} Early measured Ambisonic RIR datasets like OpenAIR \cite{murphy2010openair} and TAU-SRIR \cite{politis2022tau} were foundational but limited in scale or Ambisonic order, typically not exceeding 4th order \cite{gotz2021dataset}. To address the issue of scale, the synthetic HARP dataset introduced over 100,000 7th-order RIRs \cite{saini2025harp}. However, HARP's realism is constrained by two factors: its use of simple "shoebox" geometries, which lack ecological validity, and its reliance on the Image Source Method (ISM). While computationally efficient, ISM cannot model crucial wave phenomena like diffraction, leading to RIRs that are less accurate in complex, furnished spaces.

\textbf{Geometric and hybrid datasets:} The \textbf{SoundSpaces} platform is a notable FOA dataset: it provides approximately 17.6 million RIRs in 1st-order Ambisonics for 103 large-scale real-world indoor scenes (Matterport3D \cite{matterport} and Replica \cite{replica}) using bidirectional path tracing \cite{chen20soundspaces}. SoundSpaces is widely used for embodied audio-visual research, but is limited to FOA order and purely geometric simulation. Recently, \cite{GWA} introduced \textbf{Geometric-Wave Acoustic dataset} (\textbf{GWA dataset}) containing approximately 2 million synthetic RIRs across 6.8K furnished houses (semantically labeled 3D meshes). Crucially, GWA uses a hybrid simulation approach, low-frequency acoustic wave effects (diffraction, modal resonances) are captured by a time-domain finite-difference solver, while high frequencies are handled by ray tracing. The two are calibrated to produce accurate low- and high-frequency responses. This is the first large-scale dataset to demonstrate the benefits of combined wave-geometric acoustic modeling.

\textbf{Importance of High-Order Ambisonics:}
The choice of Ambisonics order is a trade-off between perceptual quality and computational cost. It is well-established that quality improves with order, overcoming the poor localization, diffuse sources, and timbral coloration inherent in First-Order Ambisonics \cite{BenHur2017}. While the 3rd/4th order is often considered a perceptual "sweet spot" offering robust immersion and localization \cite{Zotter2013, Schoeffler2018}, higher orders provide critical refinements for hifi applications. While some studies suggest 5th order is sufficient for high-quality headphone playback \cite{engel2021perceptual}, rendering complex sound fields over loudspeakers may require up to 7th order \cite{frank2014make}.

Our use of 7th-order Ambisonics aims for maximum perceptual fidelity. This high order raises the spatial aliasing frequency, yielding a more spectrally stable sound field for critical auralization \cite{Daniel2003}, and is necessary to render truly point-like sources, as source compactness continues to improve beyond the 3rd-order "sweet spot" \cite{Schoeffler2018}. The practical benefits of this approach are empirically validated by our Direction-of-Arrival estimation experiment in Section \ref{doa estim}.

\textbf{Gap and novelty:} Existing HOA datasets are limited by low spatial resolution (FOA-only), small scale, or purely geometric simulation. For instance, the large-scale GWA dataset \cite{GWA} is single-channel, while the 7th-order HARP dataset \cite{saini2025harp} uses a geometric-only simulation in simple shoebox rooms. HiFi-HARP fills this gap by providing the first large-scale collection of 7th-order Ambisonic RIRs generated with high-fidelity hybrid simulations in realistically furnished environments.

\begin{figure} \
    \centering
    \includegraphics[width=0.85\linewidth]{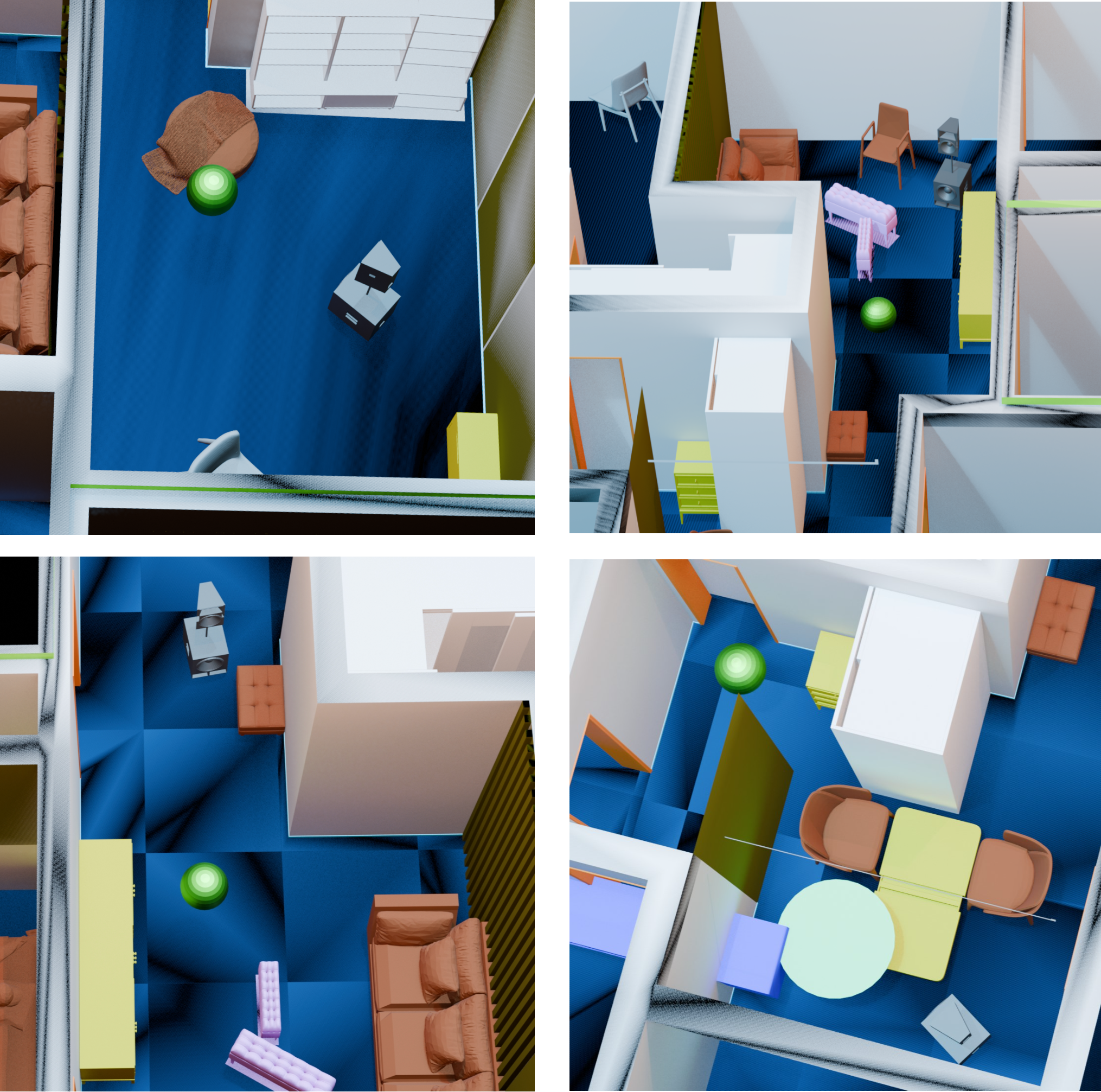}
    \caption{Example measurement configurations where green sphere is the spherical microphone array. The object colors do not accurately describe the material property and have been randomly assigned for visualisation purposes.}
    \label{fig:example_measurement}
\end{figure}

\section{Methodology: Dataset Creation}
\label{sec:method}

\subsection{3D-FRONT Scenes and Material Assignment}
HiFi-HARP is built upon the \textbf{3D-FRONT} dataset \cite{3dfront}, a vast repository of furnished indoor scenes. This dataset provides 18,968 professionally designed rooms, offering diverse geometries, furniture arrangements, and materials suitable for realistic acoustic simulation. Figure \ref{fig:example_measurement} shows a representative 3D-FRONT scene. We utilize a broad subset of these scenes, encompassing varied room volumes, shapes, and furnishing.

For acoustic simulation, each surface (walls, floors, furniture) is assigned a frequency-dependent acoustic absorption coefficient. Following \cite{GWA}, we implement a \textit{semantic material mapping}: textual labels (e.g., “wooden floor,” “tiled wall”) are matched to measured absorption spectra via a language-based classifier (SentenceFormer \cite{sbert}). This approach ensures that materials reflect real-world acoustic properties, drawing from a lookup table of publicly available absorption curves consistent with 3D-FRONT semantic labels.

\subsection{Microphone Array Design}
Our RIR simulations employ distinct microphone configurations for different frequency bands. For the low-frequency, wave-based simulation, we utilize a 64-channel spherical microphone array. The virtual microphones are arranged on the sphere's surface using a Fliege-Maier grid \cite{fliege2009integration} for near-uniform angular sampling. A sphere radius of 42 cm was chosen for two reasons: it matches the commercially available Eigenmike EM64, ensuring practical relevance, and it provides sufficiently distinct sample points for our time-domain finite-difference solver, avoiding waveform redundancy and excessive computational costs associated with smaller arrays.

For high-frequency simulations, we adopt the approach by Zang and Kong \cite{zang2025gsoundsir}, built upon the G-Sound library \cite{gsound}. This method efficiently performs ray-tracing simulations by tracking rays and storing directional data directly in the spherical harmonic domain, similar to HARP \cite{saini2025harp}, thereby bypassing the need for an explicit microphone array that would otherwise increase simulation complexity. This enables direct HOA simulation up to 9th order.

\subsection{Hybrid Acoustic Simulation}
Acoustic propagation is computed across two frequency bands. For frequencies up to 900 Hz, a \textbf{wave-based FDTD solver} is employed \cite{hamilton2021pffdtd}. This solver accurately models wave phenomena like diffraction, interference, and room modes, using a grid resolution of approximately 2 cm to represent wavelengths down to 900 Hz, with impedance boundary conditions derived from our material assignments.

For frequencies above 900 Hz, we transition to \textbf{geometric ray tracing} \cite{zang2025gsoundsir, TangImproving, gsound, schissler2017}, which efficiently handles specular reflections and scattering. The low-frequency FDTD output is converted to the spherical harmonic domain and then smoothly merged with the high-frequency ray-tracing results using a weighted crossover strategy, as detailed in \cite{GWA}. This approach ensures continuity at the 900 Hz crossover, yielding a final RIR that combines wave-theoretic accuracy for low frequencies with geometric realism for high frequencies.

Source signals are modeled as point impulses. For each house model, we typically simulate 5 sources and 10 receiver combinations, resulting in 50 measurements per room. The receiver (64-mic array center) is fixed at 1.5 m height, with source-receiver distances and orientations sampled to represent common listening scenarios.

To overcome the immense computational load of simulating 3,200 (64 mics $\times$ 5 sources $\times$ 10 receivers) solver runs per room, we developed a custom pipeline. This pipeline leverages each solver's ability to handle multiple receiver positions in a single batch. By grouping all 64 microphone locations into ten composite simulation jobs, we effectively reduce the per-room workload to just 10 runs. This optimization accelerates dataset generation by a factor of \textbf{320$\times$}, making large-scale RIR synthesis tractable.

\begin{figure}[htbp]
    \centering
    \includegraphics[width=0.8\linewidth]{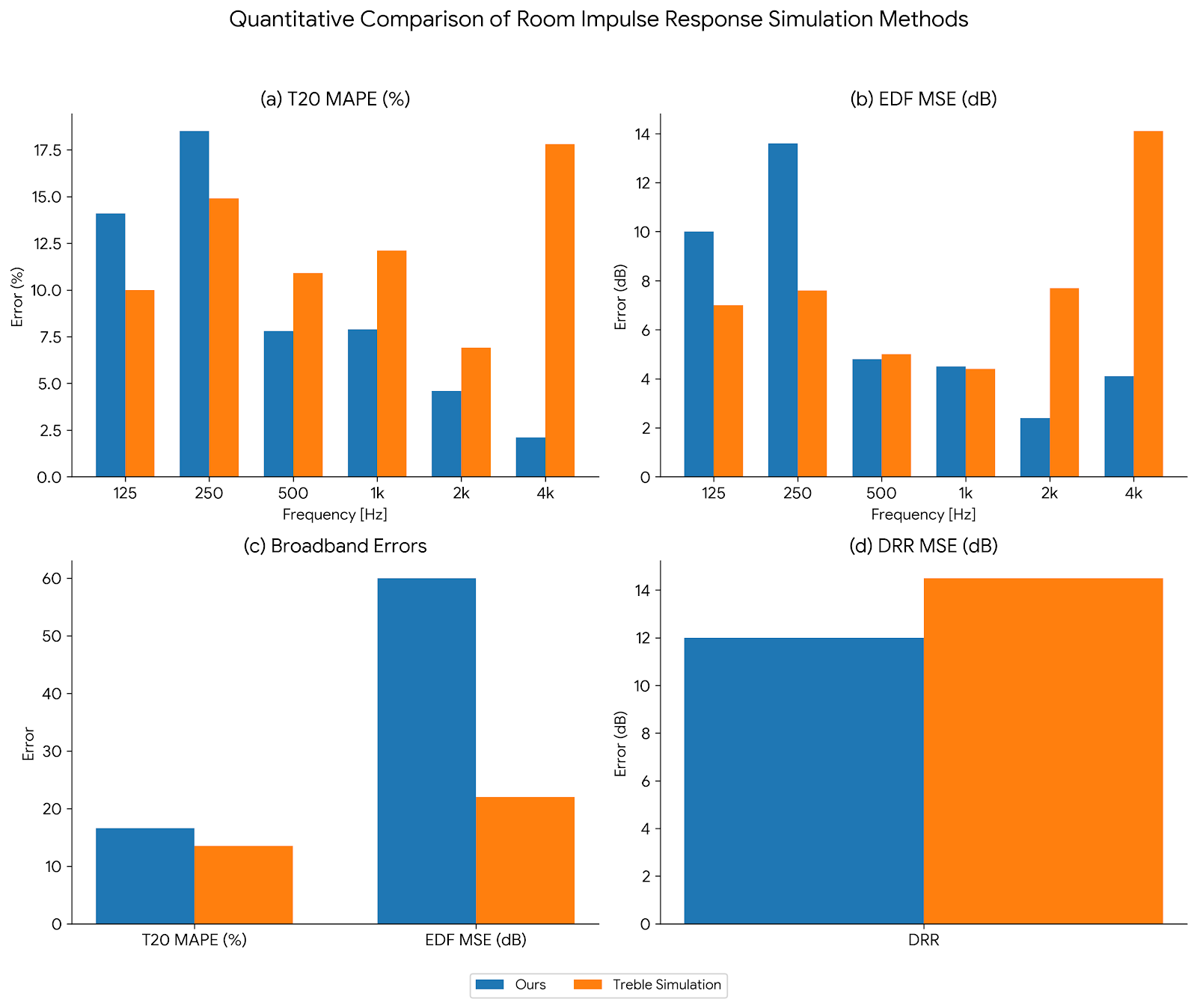}
    \caption{\textbf{Quantitative Comparison of Treble Simulation and proposed approach.} Broadband and octave-band T20 MAPE and EDF MSE, and DRR MSE of the simulated RIRs compared to the measured RIRs in room as provided in \cite{Genda}. \textbf{Note:} The differences in results are likely due to the unavailability of precise material descriptions.}
    \label{fig:treble_compare}
\end{figure}

\begin{figure} 
    \centering    
    \fbox{\includegraphics[width=0.99\linewidth]{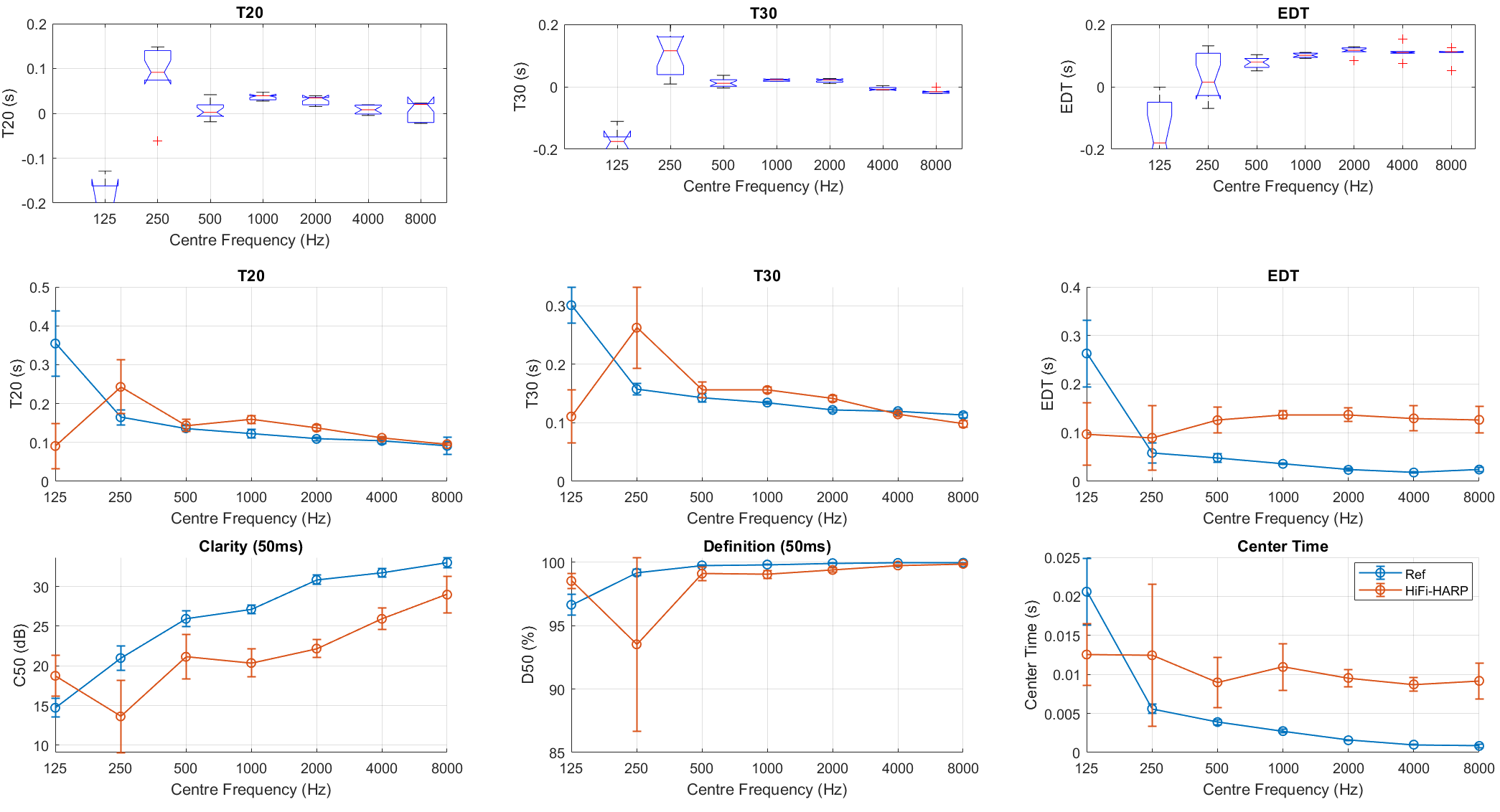}}
    \caption{Comparison of parameters extracted from real measurement done using Eigenmike EM32 and our approach. Top row demonstrates the boxplot of errors for the respective measured position.}
    \label{fig:ambilab}
\end{figure}

We validated our simulation against a commercial model (Treble) and our own lab measurements. When reproducing the GENDA Task 1 Challenge RIRs \cite{Genda}, our results diverged from the reference simulation (Fig. \ref{fig:treble_compare}). We attribute this discrepancy to our material estimation, which is based on semantic labels rather than the empirically measured coefficients used by Treble. Specifically, our method relies on textual labels (e.g., “floor” or “ceiling”), which often fail to capture the true acoustic properties of the underlying materials. In contrast, the Treble simulation employs empirically measured absorption and scattering coefficients, offering greater physical accuracy. Furthermore, comparisons with lab measurements showed minor low-frequency errors, likely from geometric simplification (Fig. \ref{fig:ambilab}). Future work could improve fidelity by adopting a multimodal approach for material and geometry estimation.

\begin{figure}
    \centering
    \includegraphics[width=0.9\linewidth]{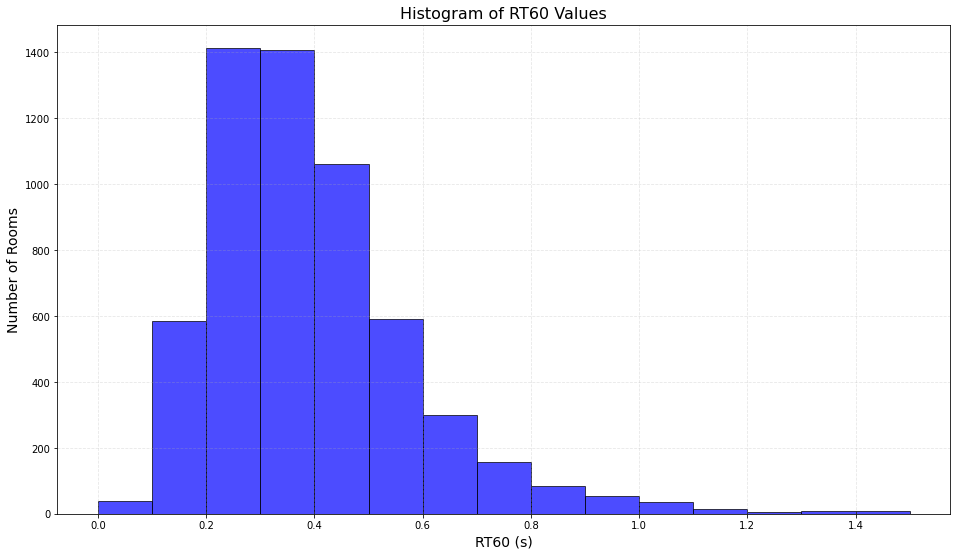}
    \caption{RT60 Distribution of the dataset}
    \label{fig:RTHistogram}
\end{figure}

\section{Dataset Statistics and Comparison }
\label{sec:Comparison}

\begin{table*}[ht]

\centering
\resizebox{0.92\textwidth}{!}{\begin{tabular}{|l |l| l| l |l|}
\toprule

\textbf{Dataset} & \textbf{Order} & \textbf{\# RIRs} & \textbf{Scenes / Variation} & \textbf{Simulation} \\
\hline
\textbf{BUT-ReverbDB \cite{Szoke_2019}} & 0th & 1300  & 11 real rooms & Measured \\
\textbf{MESH-RIR \cite{meshrir}} & 0th & 4400  & 1 room, 2D grid & Measured \\
\textbf{GWA \cite{GWA}} & 0th & 2 M & $\approx 6000$ realistic interiors & Hybrid Wave + Ray \\
\textbf{dEchorate \cite{dechorate}} & Linear Array & 1800  & variable acoustic conditions & Measured \\
\textbf{AIR \cite{brirdataset1}} & Binaural & 200+  & 4 rooms & Measured \\
\textbf{OpenAIR \cite{murphy2010openair}} & 1st & 50  & 50 extreme real rooms & Measured \\
\textbf{C4DM \cite{c4dm}}& 1st & 700 & 3 rooms, multiple positions & Measured \\
\textbf{TAU-SRIR \cite{politis2022tau}} & 1st & 114 & 9 rooms, multiple positions & Measured \\
\textbf{SoundSpaces \cite{chen20soundspaces}} & 1st & 17.6M & $\approx 100$ realistic interiors & Geometric (ray tracing) \\
\textbf{PAN-AR \cite{PANAR}} & 2nd & 21 & 4 real rooms, includes ambient noise/images & Measured HOA \\
\textbf{MOTUS \cite{gotz2021dataset}}& 3rd & 3320 & 1 room, 830 furniture arrangements & Measured \\
\textbf{ARNI-SRIR \cite{arni}} & 4th & – & 5 variable acoustic conditios, 6-DoF & Measured \\
\textbf{HOMULA-RIR \cite{HOMULA}} & HO mics + ULA & 25 & 1 seminar room, multiple positions & Measured \\

\textbf{HARP \cite{saini2025harp}} & 7th & 100K+ & Wide variety of synthetic shoebox rooms & Image-source (ISM) \\
\textbf{HiFi-HARP (ours)} & \textbf{7th} & 100K+ & $\approx 2000$ realistic interiors & Hybrid Wave + Ray\\
\bottomrule
\end{tabular}}
\vspace{2pt}
\caption{Comparison of HOA/Ambisonic RIR datasets. Entries indicate ambisonic order, scale, scene diversity, and simulation method. HiFi-HARP offers 7th-order HOA with hybrid simulation in complex indoor scenes.}
\label{table:ComparisonRIR}
\end{table*}

HiFi-HARP’s rooms span a wide range of acoustic conditions. The 3D-FRONT scenes produce room volumes from small ($\approx$30$m^2$ bedrooms) to large ($\approx$200$m^2 $ living/dining halls). We measure the omnidirectional RT60 for each RIR and only synthesize the rooms that fall under 1.2 seconds, covering most daily scenarios. Our RT60 distribution, shown in Fig. \ref{fig:RTHistogram}, is concentrated in the 0.2–0.8s range and covers typical indoor scenarios, similar to the distribution in \cite{saini2025harp}. Our material mixture (wood, carpet, glass, etc.) ensures varied absorption: on average, mid-frequency absorption coefficients range from 0.2 (highly reflective) to 0.9 (highly damped) across rooms.


\textbf{Comparison with existing datasets:} Table \ref{table:ComparisonRIR} summarizes key features of HiFi-HARP and other Spatial RIR datasets. Notably, HiFi-HARP and HARP are the only ones offering 7th-order recordings. SoundSpace provides orders only up to 1st order, albeit at a much larger scale (17.6M RIRs). Geometric simulation (ray tracing) is used in SoundSpaces, whereas HiFi-HARP improves on this by incorporating low-frequency wave effects. Smaller measured datasets (C4DM/Aalto, TAU, MOTUS, ARNI, HOMULA, PAN-AR) have at most 4th order and far fewer samples. HARP also includes approximately 100k RIRs in 7th order, but relies solely on image-source simulation without wave effects. Our HiFi-HARP complements these by simulating realistic multi-room scenes and combining wave/geometry for enhanced fidelity.

\textbf{Downstream Task Evaluation}
To demonstrate the practical value of HiFi-HARP, we evaluated its effectiveness in two representative downstream tasks. First, we tested its utility as a data augmentation resource for improving the robustness of a room acoustic parameter estimator. Second, we investigated how the high Ambisonic order of our RIRs directly impacts the generalization performance of a data-driven DOA estimator.

\subsection{Room Acoustic Parameter Estimation}
Accurate estimation of parameters like reverberation time ($T_{60}$) is a challenging blind task, benchmarked in challenges like ACE \cite{Acechallenge}. To show our dataset's value for data augmentation, we trained a model for $T_{60}$ estimation \cite{saini2024end, AudMobNet} under two conditions: 1) using a baseline training set of only measured BRIRs, and 2) using the same set augmented with 30,000 BRIRs derived from HiFi-HARP.

Table \ref{table:RTestim} shows us augmenting the training data with our synthetic RIRs significantly improved performance on a real-world test set, boosting the Pearson correlation ($\rho$) from 0.85 to 0.92. This highlights the dataset's effectiveness for improving the robustness and accuracy of acoustic estimation models.

\begin{table}
\centering
\caption{Performance of a $T_{60}$ estimation model trained with and without augmentation from the HiFi-HARP dataset.}

\resizebox{\linewidth}{!}{%
\begin{tabular}{@{}lccc@{}}
\toprule
\textbf{Training Dataset} & \textbf{Pearson Corr. ($\rho$)} ($\uparrow$) & \textbf{MSE} ($\downarrow$) & \textbf{Bias} ($\downarrow$) \\ \midrule
Measured BRIRs only & 0.85 & 0.018 & 0.01 \\
Measured BRIRs + HiFi-HARP & \textbf{0.92} & \textbf{0.012} & 0.01 \\ \bottomrule
\end{tabular}%
}
\label{table:RTestim}
\end{table}

\subsection{Direction-of-Arrival Estimation}
\label{doa estim}
Next, we investigated if the high spatial fidelity of our RIRs translates to better performance. We trained four identical 3-layer CNNs to estimate source azimuth from binaural audio. Each model was trained on 10,000 samples derived from a different Ambisonic order (1st, 3rd, 5th, and 7th) from our dataset. All models were then tested on an unseen set of real measured BRIRs \cite{brirmatching} to assess generalization.

Table \ref{table:doa_results} reveals a monotonic improvement in performance as the order of the ambisonics increases. The model trained on 7th-order-derived BRIRs achieved the lowest error, demonstrating that the spatial fidelity captured in our simulations provides a direct and measurable benefit to the performance of ML models. This validates the importance of high-order simulations for training robust spatial audio algorithms.

\begin{table}
\centering
\caption{DOA estimation performance on a real BRIR test set. Models were trained on data derived from different Ambisonic orders.}
\label{table:doa_results}
\resizebox{0.8\linewidth}{!}{%
\begin{tabular}{@{}lcc@{}}
\toprule
\textbf{Training Data} & \textbf{MSE} ($\downarrow$) & \textbf{Pearson Corr. ($\rho$)} ($\uparrow$) \\ \midrule
1st-Order ARIRs & 2.34 & 0.85 \\
3rd-Order ARIRs & 2.32 & 0.89 \\
5th-Order ARIRs & 2.26 & 0.90 \\
7th-Order ARIRs & \textbf{1.93} & \textbf{0.90} \\ 
\bottomrule
\end{tabular}

}
\end{table}

\section{Discussion }
\label{sec:discussion}

The creation of HiFi-HARP opens several avenues for research in data-driven spatial audio and acoustics. Our downstream task evaluations provide strong empirical evidence for the dataset's dual utility: the realism of the furnished scenes makes it a powerful resource for data augmentation (as seen in $T_{60}$ estimation), while the high-order spatial information directly translates to improved model generalization for spatially-aware tasks like DOA estimation. This positions the dataset as a valuable tool for training and benchmarking a new generation of algorithms in immersive audio rendering for VR/AR, cross-modal acoustic simulation (e.g., image-to-RIR), and blind acoustic scene analysis, FOA-to-HOA spatial upsampling, Spatial denoising etc.

\textbf{Limitations:} Despite its fidelity, HiFi-HARP has constraints. As with \cite{saini2025harp}, our use of deterministic simulation omits some real-world effects: fine details of furniture, the presence of humans in rooms are not modeled explicitly. Future improvements could involve regressing acoustic material parameters from 3D model context or images, rather than assigning them based on object names. Additionally, all rooms and sources are static; dynamic scenes (moving sources/receivers) are not included. The 64-channel sphere itself is an approximation of an ideal continuous array. As mentioned earlier, the size of the spherical array was kept as 42 cm resulting in spatial aliasing above 900 Hz. Finally, future work may enrich the dataset with more object types and room models.

\section{Conclusion}
\label{sec:conclusion}

We have presented HiFi-HARP, a novel dataset of high-fidelity 7th-order Ambisonic RIRs in realistic indoor scenes. By combining 3D-FRONT’s rich interiors with a hybrid wave-based and geometric simulation and a 64-channel spherical array, HiFi-HARP offers a unique resource for spatial audio research. We described the dataset pipeline, provided comparisons to existing datasets, and proposed tasks for leveraging the data. Moving forward, HiFi-HARP can support machine learning for sound field capture, acoustic inference, and immersive audio, while also highlighting the trade-offs of synthetic spatial data. We invite researchers to use HiFi-HARP to push the frontiers of 3D audio and acoustics.

\bibliography{bibliography} 
\bibliographystyle{ieeetr}

\end{document}